\documentstyle[preprint,aps,eqsecnum]{revtex}
\begin{document}

\title
{Magnetoresistance and Hall Constant of Composite Fermions}

\author{D.V. Khveshchenko}
\address
{NORDITA, Blegdamsvej 17, Copenhagen DK-2100, Denmark}

\maketitle

\begin{abstract}
\noindent
We consider both disorder and interaction effects on the magnetoresistance
and Hall constant of composite fermions in the vicinity of half
filled Landau level. By contrast to the standard case of Coulomb interacting
two-dimensional electron gas  
we find logarithmic temperature corrections to the Hall conductivity
and the magnetoresistance of composite fermions whereas 
the Hall constant acquires no such correction in the lowest order.
The theory provides a possible explanation of
the resistivity minimum at filling factor $\nu=1/2$.

\end{abstract}
\pagebreak

The theory of interaction effects in the disordered two-dimensional electron gas in a magnetic field 
developed over a decade ago \cite{AA} receives now a new boost since the discovery of compressible metal-like
states at even denominator fractions \cite{J}. 

The theory of composite fermions (CF) proposed in the seminal paper  by Halperin, Lee, and Read \cite{HLR}
offers a successful explanation of the strikingly simple Fermi liquid-like behavior observed in a surface acoustic wave
propagation \cite{W} and geometric resonance \cite{res} experiments. The impressive agreement between the observed Fermi liquid-like features and the theoretically computed electromagnetic response functions occurs already on the level of the random phase approximation (RPA).
Given the complexity of the problem and the absence of a small parameter controlling the applicability of the RPA 
(the ratio ${e^2\over v^{*}_F}$ which is typically small if one uses the value of the electron band mass $m_b \sim 0.07 m_0$ 
in $GaAs/Al_{x}Ga_{1-x}As$ becomes of the order of unity in the case of CF whose effective Fermi energy is determined by the Coulomb interaction)
such an agreement might look quite amazing. 

To some extent the situation was clarified in a number of more recent publications \cite{SH},\cite{AIM},\cite{KLW}
 where it was shown that despite of the strong
gauge interactions between CF the (gauge invariant) physical response functions manifest almost no deviation from their Fermi liquid
counterparts in the low-energy long-wavelength regime relevant for the description of the experiments \cite{W},\cite{res}.
Even though there is no parameter which justifies the RPA 
the similarity to the Fermi liquid response stems from asymptotic low-energy Ward identities based on the underlying particle number conservation law.

On the other hand, the behavior of the single particle Green function is known to change drastically by the long-range
retarded gauge interactions. And so do the observables which can be related to the gauge-invariant part of the single CF propagator. 
One example of this sort is the behavior of magnetoresistance (MR) $\rho_{xx} (B)$ away from $\nu=1/2$ 
which demonstrates oscillations reminiscent of the Shubnikov-de Haas (SdH) effect at low fields. 

Such oscillations develop at large enough deviations from half filling $\Omega_c ={\Delta B \over m^*}>1/\tau_{tr}$ where 
$\Delta B = B-4\pi n_{e}$ and $1/\tau_{tr}$ is an elastic scattering rate of CF with the density $n_e$ and the effective mass
$m^* \sim {k_F\over e^2}$. In the regime $\Omega_c \tau_{tr} >>1$ one can calculate the oscillating part of the MR
$\Delta\rho_{xx} (B)=\sum_{k}I_k
\cos{4\pi^2 n_e\over {\Delta B}}$ semiclassically 
by relating the amplitude of the $k^{th}$ harmonics $I_k$ to the phase factor of the CF making $k$ laps along the closed 
cyclotron
orbit.  The corresponding Dingle plot exhibits a magnetic field 
dependent effective mass $m^{*}(\Delta B)$ which has a tendency to diverge in the $\Delta B\rightarrow 0$ limit.
The divergency $m^{*}(\Delta B)\sim \log \Delta B$ was recently found \cite{LMS} to be
in agreement with earlier perturbative calculations of the energy-dependent effective mass 
  carried out in the case of $T=0$ and no
impurities \cite{HLR},\cite{SH}.

However, so far the theory did not succeed in reconciling with the experiment which shows a much stronger divergency
$m^{*}(\Delta B)\sim (\Delta B)^{-4}$  \cite{Du} (in the case of a 2D hole system an even stronger dependence $\log m^* (\Delta B)\sim (\Delta B)^{-3/2}$ was reported in \cite{M}).

 Such a discrepancy is attributed to the effect of disorder which is barely taken into account in the present theory and which is believed to become dominant in the vicinity of $\nu=1/2$. Considering the effect of disorder 
but neglecting the gauge interactions  the authors of the Ref.\cite{MAW} managed to reproduce the observed scaling
of the Dingle plot  enveloping function  with the effective field: $\log {\Delta\rho_{xx} (\Delta B)\over \rho_xx}\sim (\Delta B)^{-4}$ \cite{C}. However, the naive estimate of the coefficient 
appears to be about 2300 times greater than the measured one. Such a drastic conflict with the 
experiment implies a necessity of a more accurate account of disorder and/or gauge interactions of CF.

In the present letter we study combined disorder and interaction effects on the MR in the regime of
 low effective fields
($\Omega_c ={\Delta B \over m^*}\leq 1/\tau_{tr}$) preceding the onset of the SdH oscillations.

On theoretical side, one obvious challenge is to explain a broad minimum of $\rho_{xx} (B)$ at $\nu=1/2$
which  suggests a positive MR in the CF theory. The previous studies addressing this issue also ignored CF
gauge interactions and concentrated on the effect of Coulomb impurities (ionized donors) which become sources of the static random gauge magnetic field (RMF) after the mapping of electrons to the CF \cite{HLR},\cite{KZ}.

Since the RMF brakes time-reversal symmetry there is no Cooperon pole in the particle-particle channel of the two-particle Green function \cite{KZ} which eliminates conventional weak localization logarithmic temperature corrections and
the related effect of negative MR in the external magnetic field. 
On the basis of this argument and the numerical simulations of the lattice version of
the RMF problem it was suggested in \cite{AWKZ} that by contrast to the case
of ordinary potential disorder the RMF scattering results in the positive MR. 

On the other hand, a semiclassical analysis of the RMF problem carried out in \cite{DV2} by solving the Boltzmann
equation beyond the relaxation time approximation reveals a non-trivial negative MR
\begin{equation}
{\Delta\rho_{xx}\over \rho_{xx}}|_{RMF}= - 0.06 (\Omega_c \tau_{tr})^2
\end{equation}
The result (1) was obtained in \cite{DV2} for the 
case of spinless fermions with isotropic dispersion and circular Fermi surface and a long-range correlated RMF
described by the correlation function $<b_q b_{-q}>=2\pi^2 \alpha n_e e^{-2q\xi}$ provided that $k_F \xi >>1$. This case which, apparently,  is hardly
accessible in numerical simulations describes the $\nu\sim 1/\Phi$ problem at $\alpha={1\over 2}\Phi^2$ and $\xi$ equal to the width
of the spacer between the donor layer and the 2D electron gas (typically,  $k_F \xi \sim 15$ \cite{HLR}). 
It remains to be seen, however, whether or not the semiclassically computed negative MR (1) is related to the localization 
phenomena in the RMF.

It was also argued in \cite{DV2} that one can only justify the use of the Born approximation in the RMF problem at small $\alpha$.
In the case of interest ($\alpha=2$  for $\nu=1/2$ or even larger for higher even denominator fractions)  the Born approximation
systematically overestimates the strength of the RMF scattering. However, provided $k_F \xi >>\sqrt\alpha$ one can resort
on the eikonal-type solution of the Boltzmann equation which gives the CF transport time and the classical conductivity  
\begin{equation}
\tau_{tr}={\sigma_{xx}\over \epsilon_F}{h\over e^2}={2\xi\over v_F} e^{\alpha}K_{1}(\alpha)\approx {2\xi\over v_F}
\end{equation}
The result (2) is about twice the value of the Born estimate \cite{HLR} which provides an essentially better agreement  
with the measured  $\rho_{xx}$ at $\nu=1/2$ at not very low temperatures. A further improvement can be, presumably, achieved by taking into account correlations in positions of charged impurities.

The complete theory  should, however, include the CF gauge interactions as follows, for example, 
from the strong sample-dependent logarithmic temperature correction to $\rho_{xx}$ at $\nu=1/2$ observed in the temperature range from 0.5K to 15 mK \cite{log}. A possible explanation of the $\log T$ term as resulting from the first order gauge interaction correction in presence of disorder was proposed in \cite{DV1}. 

In the Coulomb gauge ${div}{\vec A}=0$ the gauge interactions of CF are described 
by the propagator which 
has the following form in the disordered regime $\omega\tau_{tr} <<1$ and $ql<<1$ ($l=v_F\tau_{tr}$):
\begin{equation}
D^{-1}_{\mu\nu}(\omega, q)=
\pmatrix
{
N(\epsilon_F){Dq^2\over {Dq^2 - i\omega}} &
-i{q\over 4\pi}
\cr
i{q\over 4\pi} & 
-i N(\epsilon_F)D\omega + \chi_{q}q^2 \cr}
\end{equation}
Here $D={1\over 2}v^{2}_{F}\tau_{tr}$ is the CF diffusion coefficient,
$N(\epsilon_F)$ is the CF density of states on the Fermi level,
 and $\chi_q ={1\over 12\pi m^*}+{1\over (4\pi)^2}V_q$ is the effective orbital magnetic susceptibility determined 
by  the pairwise electron potential $V_q$.

Since the electron Coulomb interaction $V_q ={2\pi e^2\over q}$
can be screened by placing a ground plate close to the 2D electron gas it is worthwhile to consider both cases of the Coulomb and the short-range
$(V_q \approx V_0 ={2\pi e^2\over \kappa}$ where $\kappa$ is
a screening constant) potentials.
 
It was shown in \cite{DV1} that the leading negative
logarithmic temperature correction to $\Delta\sigma_{xx}$ is due to the transverse component 
$D_{11}(\omega, q)$. It is enhanced by the factor $\log (k_F l)$ as compared to the 
contribution coming from the scalar component $D_{00}(\omega, q)$ which coincides with the well-known result at zero field \cite{AA},\cite{LR}.
In the short-range case we obtained
\begin{equation}
\Delta\sigma^{CF}_{xx}(\Delta B=0)={e^2\over 2\pi h}(\log {T\tau_{tr}}) \log ({k_F l})
\end{equation}
whereas in the case of the unscreened Coulomb potential double-logarithmic terms appeared 
\begin{equation}
\Delta\sigma^{CF}_{xx}(\Delta B=0)={e^2\over 2\pi h}(\log {T\tau_{tr}}) 
[\log ({k_F l})+{1\over 4}\log {T\tau_{tr}}]
\end{equation}
which reduce the correction (5) with respect  to (4) by a factor of two in the range of temperatures
$T_0 =\epsilon_F {1\over (k_F l)^3} <T<1/\tau_{tr}$ ( at $T<T_{0}$ the divergency in (5) is cut off). 
The logarithmic corrections are non-universal (they are stronger in cleaner samples of higher density) but
only weakly dependent on
details of the interaction potential $V_q$.

It turns out that in presence of a finite effective field $\Delta B$ similar corrections to the CF Hall conductivity occur. 

In the case of ordinary electrons logarithmic temperature corrections to $\sigma_{xy}$ are known to appear only in the localization theory whereas the Coulomb interaction alone does not produce such terms \cite{LR}. The latter result was originally obtained in the weak field limit and in the first order in the RPA-screened Coulomb interaction \cite{AKLL}
and then extended (within the same treatment of the interaction) 
to high fields $\Omega_c \tau_{tr}\geq 1$ including the quantum limit $\Omega_c \sim \epsilon_F$ \cite{HSY}.

It also follows from the analysis \cite{HSY} that the correction $\Delta\sigma_{xx}=
{ e^2\over \pi h}\log {T\tau_{tr}}$ weakly depends on $B$ via the effective transport time
 $\tau_{tr}(B)$ and leads to a positive contribution to the MR which reads as ${\Delta\rho_{xx}\over \rho_{xx}}
\sim {1\over k_Fl}\log \Omega_c\tau_{tr} $ at $\Omega_c\tau_{tr}>>1$ when $\tau_{tr}(B)\approx ({\pi\tau_{tr}/2\Omega_c})^{1/2}$.
This effect should not be confused with a well-known positive MR resulting from the spin-triplet channel of the Hartree
term \cite{LR} which is apparently absent in the spin-polarized case.

Nevertheless, the overall 
MR of spin-polarized electrons
appears to be negative due to the Drude-type $B$-dependence of the bare (classical) conductivity.
In the general case, when both $\sigma_{xx}$ and $\sigma_{xy}=(\Omega_c\tau_{tr})\sigma_{xx}$ 
 acquire some Coulomb exchange corrections
$\Delta\sigma_{xx}(B)$ and $\Delta\sigma_{xy}(B)\sim B$ the low-field MR can be cast in the form 
\begin{equation}
{\Delta\rho_{xx}\over \rho_{xx}}=({\Delta\sigma_{xx}(0)\over \sigma_{xx}}-{\Delta\sigma_{xy}\over \sigma_{xy}})
(\Omega_c \tau_{tr})^2 -({\Delta\sigma_{xx}(B)\over \sigma_{xx}}-{\Delta\sigma_{xx}(0)\over \sigma_{xx}})
\end{equation}
Therefore, in the case when localization effects are suppressed $( \Delta\sigma_{xy}=0 )$  the MR is governed
by the negative correction $\Delta\sigma_{xx}(0)$  and behaves as ${\Delta\rho_{xx}\over \rho_{xx}}\sim 
(\Omega_c \tau_{tr})^2 {1\over k_F l}\log T\tau_{tr}$. 

We notice, by passing, that in the non-interacting localization theory $\Delta\sigma_{xy}(B)=2(\Omega_c \tau_{tr})\Delta\sigma_{xx}(0)$
and the above terms yield a positive effect which is, however, completely dominated by the much larger
negative MR coming from the 
field dependence of $\Delta\sigma_{xx}(B)$ itself \cite{LR}.

The above conclusions are based on the assumption that at $B=0$ the Hall conductivity is zero. In the case of CF this assumption was recently questioned in \cite{D-H} where it was argued that due to the time-reversal asymmetric 
Chern-Simons interactions 
 $\sigma^{CF}_{xy}$ is finite. After an initial confusion the authors of \cite{D-H}
admitted that no such contribution occurs in the first order in $D_{01}(\omega, q)$. They argued, however, that the predicted
value $\sigma^{CF}_{xy}=-{e^2\over 2h}$ follows from such general principles as a particle-hole symmetry at $\nu=1/2$. 

Although the question remains to be open for now, one might think that even if such a bare $\sigma^{CF}_{xy}$ were present
in the theory it had to be considered as a high-frequency ($1/\tau_{tr}<<\omega << \epsilon_F$) value of $\sigma^{CF}_{xy}(\omega)$
which should not appear in the zero frequency response of the disordered system.
Otherwise, one would obtain a non-analytic (V-shaped) MR ${\Delta\rho_{xx}\over \rho_{xx}}\sim |\Delta B|$
as can be readily seen by repeating the derivation of (6) in
the case of $\sigma_{xy}(0)\neq 0$.

The calculation of the CF conductivity tensor which we sketch below yields  
\begin{equation}
\Delta\sigma^{CF}_{xx}(\Delta B)= (1-(\Omega_c\tau_{tr})^2)\Delta\sigma^{CF}_{xx}(0), ~~~~~~~
\Delta\sigma^{CF}_{xy}(\Delta B)= 2(\Omega_c\tau_{tr}) \Delta\sigma^{CF}_{xx}(0)
\end{equation}
where $\Delta\sigma^{CF}_{xx}(0)$ is given by the Eq.(4-5).

The MR of CF due to the combined effect of disorder and gauge interactions appears to be positive (we keep only the term linear in $\Delta ={\Delta\sigma^{CF}_{xx}(0)\over \sigma^{CF}_{xx}}$ but
$\Omega_c\tau_{tr}$ can be arbitrary):
\begin{equation}
{\Delta\rho^{CF}_{xx}\over \rho^{CF}_{xx}}=-\Delta (1+(\Omega_c \tau_{tr})^2)^2
\end{equation}
and its variation with $\Delta B$ is greater than the RMF contribution (1) at all temperatures below $\sim 2K$.

On the contrary, the correction to the CF Hall constant $R^{CF}_H ={\rho^{CF}_{xy}\over \Delta B}$ appears to be quadratic
in $\Delta$:
\begin{equation}
{\Delta R^{CF}_{H}\over R^{CF}_H}=
-{\Delta}^2 (1+(\Omega_c \tau_{tr})^2)^3
\end{equation}
 Provided that the physical resistivity is simply equal to $\rho^{CF}_{xx}(\Delta B)$ \cite{HLR}
we conclude that $\rho_{xx}(B)$ must exhibit a local minumum in the vicinity of $\nu=1/2$:
\begin{equation}
{\rho_{xx}(B)-\rho_{xx}(4\pi n_e)\over \rho_{xx}(4\pi n_e)}=-\Delta (\Omega_c \tau_{tr})^2
(2+(\Omega_c \tau_{tr})^2)
\end{equation}  
On the basis of the RPA-type relation between  $\rho^{CF}_{xy} (\Delta B)$ and the physical Hall resistivity $\rho_{xy}(B)=\rho^{CF}_{xy}(\Delta B) +2{h\over e^2}$ \cite{HLR} we predict
that in the first order in $\Delta$ there is no $\log T$ correction to the slope of $\rho_{xy}(B)-\rho_{xy}(4\pi n_e)\sim \Delta B$.

To comment on the consistency of our results with the high field analysis of the first order Coulomb exchange correction
carried out in \cite{HSY}
we note that the CF representation  provides an essentially  better account of the Coulomb
interaction than the conventional RPA in the original electron picture. The very cornerstone of the concept of CF,
the formation of the Fermi surface at $\nu=1/2$, results from a minimization of the Coulomb energy.
We emphasize, however, that formulae (7) originate from the first order correction in the CF transverse gauge interaction and
it is not clear at this point if higher order corrections could alter the above $\Delta B$-dependence.

Now we are going to outline the calculation which employes the method of \cite{HSY} and leads to the Eq.(7). 

A straightforward analysis shows that the first logarithmic temperature correction to $\sigma_{\alpha\beta}$ comes from
the standard exchange diagrams in Fig.(1) similar to those in Fig.4 (d,e) from the first reference in \cite{HSY}
where the wavy line now denotes the
transverse gauge (current-current) interaction  $D_{11}(\nu, {q})=
{1\over {-iN(\epsilon_F)D_H\nu  + {\chi^{\prime}_q} q^2}}$.

By contrast to the case of Coulomb interacting electrons \cite{AA},\cite{AKLL},\cite{HSY}
the diagrams in Fig.1 contain only one impurity ladder in the particle-hole channel (diffusion)
$\Gamma(\epsilon -i0, \epsilon+\nu +i0, q)={1\over {2\pi N(\epsilon_F)\tau^2}}
{1\over {D_H q^2 -i\nu}}$  where $D_H ={v^2_F\tau_{tr}\over 2}{1\over {1+(\Omega_c \tau_{tr})^2}}$ is the diffusion coefficient in a magnetic field. 

As opposed to the usual case of the 
density-density coupling the impurity dressing of current interaction vertices is purely multilplicative:
${{\vec p}\over m^*}\rightarrow{\vec \Lambda}(\epsilon -i0, \epsilon+\nu +i0)={{\vec p}\over m^*}{\tau_{tr}\over \tau}$ where $1/\tau$ stands for a total single particle scattering rate due
to disorder.  Although in the RMF problem  $1/\tau$ is formally divergent \cite{AMW} it drops out of all physical
observables which only depend on the transport time $\tau_{tr}$ given by (2). Therefore one can operate with  $1/\tau$ as if it were finite \cite{AMW}. 

We use the basis of Landau eigenfunctions and first compute the sum over fermionic states
which factorizes into two parts on either side of the impurity ladder. Each part contains two current matrix elements 
and  equals to either 
$M_{xx}=M_{yy}={i 2\pi N(\epsilon_F)\epsilon_F\tau^{2}_{tr}\over {m^{*}(1+(\Omega_c \tau_{tr})^2)}}$ 
or $M_{xy}=-M_{yx}=(\Omega_c \tau_{tr})M_{xx}$, the signs depending on the
frequencies $\epsilon$
and $\nu$ carried by 
the fermionic and the gauge lines. If the external frequency $\omega >0$  both
diagrams do not cancel out only in the domain $\epsilon <-\omega < \nu +\epsilon <0$.

 Putting all factors together and accounting for the mirror diagrams with the gauge interaction
 dressing of the particle and the hole line interchanged we arrive at the formula  
\begin{equation}
\Delta\sigma^{CF}_{\alpha\beta}(\omega)=
{ie^2\over 2\pi}
\int^{1/\tau_{tr}}_{\omega}
{d\nu\over 2\pi}
\int{d{\vec q}\over (2\pi)^2} 
{M_{\alpha\gamma}M_{\delta\beta}(\delta_{\gamma\delta}-{q_\gamma q_\delta\over q^2})
\over
 {2\pi N(\epsilon_F)\tau^{2}_{tr}(D_H q^2 -i(\nu+\omega))
(iN(\epsilon_F)D_H\nu  - {\chi^{\prime}_q} q^2 ) } }
\end{equation}
where ${\chi^{\prime}_q} = \chi_q + {1\over 8\pi m^*}$. 
At finite temperature $T>>\Omega$ we calculate (11) 
by means of the analytic continuation from imaginary frequencies.
Estimating (11) at $k_F l>>1$ and $T<<1/\tau_{tr}$ 
with logarithmic accuracy we obtain (7). 
Notice that the total RMF scattering time $\tau$ as well as $N(\epsilon_F)$ drop out of (11).
 
 Because of a singular behavior of $D_{11}(0,q)$ the leading logarithms
in (4-5) are not affected by the corresponding Hartree terms \cite{HSY},\cite{LR}.

At strong enough effective fields  the effective transport time $\tau_{tr}(B)$ and the
logarithmic enhancement factor $\log (m D_{H})$ 
become magnetic field dependent. Although it only changes the argument under the logarithm 
we do not expect our main results  (7) to remain reliable at $\Omega_c\tau_{tr}\geq 1$.

In conclusion, 
we consider the interference of disorder and interaction effects on the magnetoresistance and Hall constant of composite fermions in the vicinity of half
filled Landau level. By contrast to the case of the Coulomb interacting
 2D electron gas 
we find the logarithmic temperature correction to the Hall conductivity
 of composite fermions which leads to their positive magnetoresistance at $\Delta B \leq 1T$. 
We propose a possible explanation of the minimum of longitudinal resistivity 
and predict an absence of the leading logarithmic correction to the slope of the measured Hall resistivity at filling factor $\nu=1/2$.

The author is indebted to B.I.Halperin, P.A.Lee, P.Wolfle, A.D.Mirlin,
A.L.Rokhinson, and  G.Kotliar for valuable discussions of these
and related issues and to S.M.Girvin for pointing him out the Ref.\cite{HSY}.

\pagebreak

\end{document}